\documentclass[showpacs,preprintnumbers,english,amsmath,amssymb]{revtex4}
\usepackage[T1]{fontenc}
\usepackage[latin1]{inputenc}
\usepackage{graphicx}
\usepackage{amssymb}
\usepackage{verbatim}
\usepackage{epic,eepic}
\usepackage{babel}
\usepackage{color}

\begin{document}
\def\Journal#1#2#3#4{{#1} {\bf{#2}}, {#3} (#4).}
\def\ANP{Adv. Nucl. Phys.}
\def\ARNPS{Ann. Rev. Nucl. Part. Sci.}
\def\ADEP{Advances in High Energy Physics}
\def\CTP{Commun. Theor. Phys.}
\def\EPJA{Eur. Phys. J. A}
\def\EPJC{Eur. Phys. J. C}
\def\IJMPA{International Journal of Modern Physics A}
\def\IJMPE{International Journal of Modern Physics E}
\def\JCHP{J. Chem. Phys.}
\def\JCP{Journal of Computational Physics}
\def\JHEP{JHEP}
\def\JPCS{Journal of Physics: Conference Series}
\def\JPG{J. Phys. G: Nucl. Part. Phys.}
\def\NATURE{Nature}
\def\NC{La Rivista del Nuovo Cimento}
\def\NCA{IL Nuovo Cimento A}
\def\NPA{Nucl. Phys. A}
\def\NST{Nuclear Science and Techniques}
\def\PA{Physica A}
\def\PAN{Physics of Atomic Nuclei}
\def\PHY{Physics}
\def\PRA{Phys. Rev. A}
\def\PRC{Phys. Rev. C}
\def\PRD{Phys. Rev. D}
\def\PLA{Phys. Lett. A}
\def\PLB{Phys. Lett. B}
\def\PLD{Phys. Lett. D}
\def\PRL{Phys. Rev. Lett.}
\def\PL{Phys. Lett.}
\def\PREV{Phys. Rev.}
\def\PREP{\em Physics Reports}
\def\PROG{Progress in Particle and Nuclear Physics}
\def\RPP{Rep. Prog. Phys.}
\def\RDNC{Rivista del Nuovo Cimento}
\def\RMP{Rev. Mod. Phys}
\def\SCIENCE{Science}
\def\ZPA{Z. Phys. A.}

\def\ANN{Ann. Rev. Nucl. Part. Sci.}
\def\ANNAST{Ann. Rev. Astron. Astrophys.}
\def\AP{Ann. Phys}
\def\APJ{Astrophysical Journal}
\def\APJS{Astrophys. J. Suppl. Ser.}
\def\EJP{Eur. J. Phys.}
\def\LANC{Lettere Al Nuovo Cimento}
\def\NCA{Nuovo Cimento A}
\def\PHYS{Physica}
\def\NP{Nucl. Phys}
\def\MATH{J. Math. Phys.}
\def\JPAM{J. Phys. A: Math. Gen.}
\def\PRO{Prog. Theor. Phys.}
\def\NPB{Nucl. Phys. B}

\title{A new two-component model for hadron production in heavy-ion collisions}

\author{Xuejiao Yin$^{1}$, Lilin Zhu$^{1}$ and Hua Zheng$^{2}$\footnote{Email address: zheng@lns.infn.it}}
\affiliation{
$^{1}$College of Physical Science and Technology, Sichuan University, Chengdu 610064, People's Republic of China; \\
$^{2}$Laboratori Nazionali del Sud, INFN, via Santa Sofia, 62, 95123 Catania, Italy.}


\begin{abstract}
    Using the experimental data from the ALICE program on the centrality dependence of the transverse momentum ($p_T$) spectra in Pb+Pb collisions at $\sqrt{s_{NN}}=2.76$ TeV, we show that the double-Tsallis distribution and the generalized Fokker-Plank (FP) solution can not describe the spectra of pions, kaons and protons from central to peripheral collisions in the entire $p_T$ region, simultaneously. Hence, a new two-component distribution, which is a hydrodynamic extension of the generalized FP solution accounting for the collective motion effect in heavy-ion collisions, is proposed in order to reproduce all the particle spectra. Our results suggest that the particle production dynamics may be different for different particles, especially at very low $p_T$ region.
\end{abstract}


\maketitle

\section{Introduction}
The advent of a new generation of high energy collider experiments, such as Relativistic Heavy Ion Collider (RHIC) and Large Hadron Collider (LHC), has launched a new era in the study of the hadron production. Plenty of pp, pA and AA collisions data have been accumulated, which allow us to study the nature of the final particle production. The transverse momentum spectra carry important information about the dynamics of particle production and evolution process of interacting system formed in high energy nuclear collisions. In the past decade, the attempt to understand the particle production mechanism by different theoretical and phenomenological approaches has been a great success ~\cite{pbm, kh, gjs, pq1, pq2, jet, vg, rf, hy1,Bylinkin:2014aba, Bylinkin:2015msa,wongprd,Rybczynski:2014ura,Cleymans:2011in, Aamodt:2011zj, liuAuAu2014,Zheng:2015tua,Zheng:2015gaa, Wei:2014hsa, Liu:2008am, Liu:2008ar}. Generally, the theoretical investigation of hadron production in heavy-ion collisions is operated into different camps, characterized by the regions of transverse momenta $p_T$ of the produced hadrons. At low $p_T$ statistical hadronization and hydrodynamical models are generally adopted \cite{pbm, kh, gjs}, whereas at high $p_T$ jet production and parton fragmentation with suitable consideration of medium effects in perturbative QCD (pQCD) are the central themes \cite{pq1, pq2, jet}. The approaches have been studied essentially independent of each other with credible success in interpreting the experimental data for different $p_T$ regions, since their dynamics are decoupled. 

At intermediate and lower $p_T$ recombination or coalescence subprocess (ReCo) in heavy-ion collisions has been found to be more relevant, which has successfully explained various experimental data \cite{vg, rf, hy1}. Beside the ReCo model, there are also other phenomenological models proposed to describe the hadron production. In Refs. \cite{Bylinkin:2014aba, Bylinkin:2015msa}, a Two-Component model, the particle spectra could be treated as a summation of an exponential (Boltzmann-like) and a power-law distributions, was suggested. But this model could not describe the charged particle production at very high $p_T$ in central Pb+Pb collisions at 2.76 TeV. So an additional power-law term was added, which was explained by the peculiar shape of the nuclear modification factor $R_{AA}$.  On the other hand, due to the effect of the collective motion in large colliding system, the relativistic hydrodynamics is usually adopted to consider the particle production, instead of thermodynamic methods. Therefore, the charged particle spectra could be consisted of a hydrodynamic term and two power-law terms suggested as well in Ref. \cite{Bylinkin:2014aba}, 
\begin{eqnarray}
\frac{\mathrm{d}{N}}{p_T \mathrm{d}p_T}=A_e \int_{0}^{R}r\mathrm{d}rm_TI_0(\frac{p_T\sinh\rho}{T_e})
K_1(\frac{m_T\cosh\rho}{T_e})+
\frac{A}{(1+\frac{p_T^2}{T^2\cdot N})^N}+\frac{A_1}{(1+\frac{p_T^2}{T_1^2\cdot N_1})^{N_1}},
\label{formulaab}
\end{eqnarray}
where $\rho=\tanh^{-1}\beta_r$ is the transverse flow rapidity and the radial flow velocity is parametrized as $\beta_r(r)=\beta_s r/R$ with $\beta_s=0.5c$ for the surface velocity. $R$ is a parameter related to the transverse size of the particle distribution in space and $r$ is the distance of the particle from the origin of the coordinate system in the transverse plane. $m_T=\sqrt{m^2+p_T^2}$ is the transverse mass and $m$ is the rest mass of particle.  $I_0$ and $K_1$ are the modified Bessel functions. Hence, there are eight free parameters $A_e$, $T_e$, $A$, $T$, $N$, $A_1$, $T_1$, $N_1$, which add the difficulty to fit, even it can describe the charged particle spectra very well in central Pb+Pb collsions. Other forms of multicomponent models, which were derived from multisource thermal model  \cite{Wei:2014hsa, Liu:2008am, Liu:2008ar}, were also applied to the particle transverse momentum spectra produced from low energy to high energy heavy-ion collisions.

Besides, the Tsallis distribution, which was proposed about three decades ago \cite{Tsallis:1987eu}, has been widely applied to describe final particle production with great success by the theorists and experimentalists \cite{wongprd,Rybczynski:2014ura,Cleymans:2011in, Aamodt:2011zj, liuAuAu2014,Zheng:2015tua,Zheng:2015gaa}. It was derived in the framework of non-extensive thermodynamics,
\begin{eqnarray}
f(E, q)=A\left[1+(q-1)\frac{E-\mu}{T}\right]^{-\frac{1}{q-1}},
\label{Tsallis}
\end{eqnarray}
where $q$ is the entropic factor, which measures the nonadditivity of the entropy and $T$ is the temperature. The two parameters carry important information of the observed colliding system. $\mu$ is the chemical potential which could be assumed to be 0, when the colliding energy is high enough and the chemical potential is much smaller than the temperature. If the self-consistent thermodynamical description is taken into account, the effective distribution $[f(E, q)]^q$ is needed \cite{Rybczynski:2014ura, Cleymans:2011in, liuAuAu2014}. One should bear in mind, even though different versions of Tsallis distribution were adopted by different groups in the literature, it has established the excellent ability to describe the hadron spectra in a large $p_T$ region in pp, pA and AA collisions. Here we will try to pursue this approach to the identified particles in the non-central Pb+Pb collisions at $\sqrt{s_{NN}}=2.76$ TeV. Furthermore, for comparison, we will also discuss the particle spectra in the frame of Fokker-Planck equation and try to get some information about the hadron production during the evolution of the colliding system. 

The paper is organized as follows. In Sections \ref{Tsallis} and \ref{Fokker}, we show our results of particle spectra from Pb+Pb collisions by a double-Tsallis distribution as well as the generalized Fokker-Plank solution, respectively. A new two-component distribution which is consisted of the hydrodynamic term and generalized Fokker-Plank solution is proposed in Section~\ref{hydro}. In Section~\ref{results}, a detailed comparison among the three distributions is shown. Finally, a summary is given in Section \ref{summary}.

\section{A double-Tsallis distribution}\label{Tsallis}
In our earlier work  \cite{Zheng:2015gaa}, it has been demonstrated that a single Tsallis distribution could not fully reproduce the whole structure of the observed particle spectra in central Pb+Pb collisions at $\sqrt{s_{NN}}=2.76$ TeV. Therefore, a double-Tsallis distribution could be proposed, 
\begin{eqnarray}
\frac{\mathrm{d}N}{p_T \mathrm{d}p_T}=C_1\cdot (1+ \frac{E_T}{m_1 T_1})^{-m_1}+C_2\cdot (1+ \frac{E_T}{m_2 T_2})^{-m_2},
\label{formula1}
\end{eqnarray}
where $E_T=\sqrt{m^2+p_T^2}-m$ is the transverse energy. When the rest mass of particle $m\rightarrow 0$, (\ref{formula1}) becomes the same as the double-Tsallis distribution proposed in Ref. \cite{Rybczynski:2014ura} for charged particles, which are dominated by pions. But when $m$ is large, such as kaons, protons and antiprotons, etc., the mass effect should be taken into account. Compared with the single Tsallis distribution, three more parameters are increased, which allowed one to fit the charged particle spectra with $p_T$ up to 100 GeV/c in the most central Pb+Pb collisions \cite{Rybczynski:2014ura}. Recently,  the transverse momentum spectra of charged pions, kaons and protons up to $p_T=20$ GeV/c have been measured in Pb+Pb collisions at $\sqrt{s_{NN}}=2.76$ TeV using the ALICE detector for six different centrality classes \cite {Adam:2015kca}. Hence, it is right time for us to investigate whether the double-Tsallis distribution can describe the identified particle production both at central and non-central collisions. 

\begin{figure*}
        \centering
        \begin{tabular}{ccc}
        \includegraphics[width=0.32\textwidth]{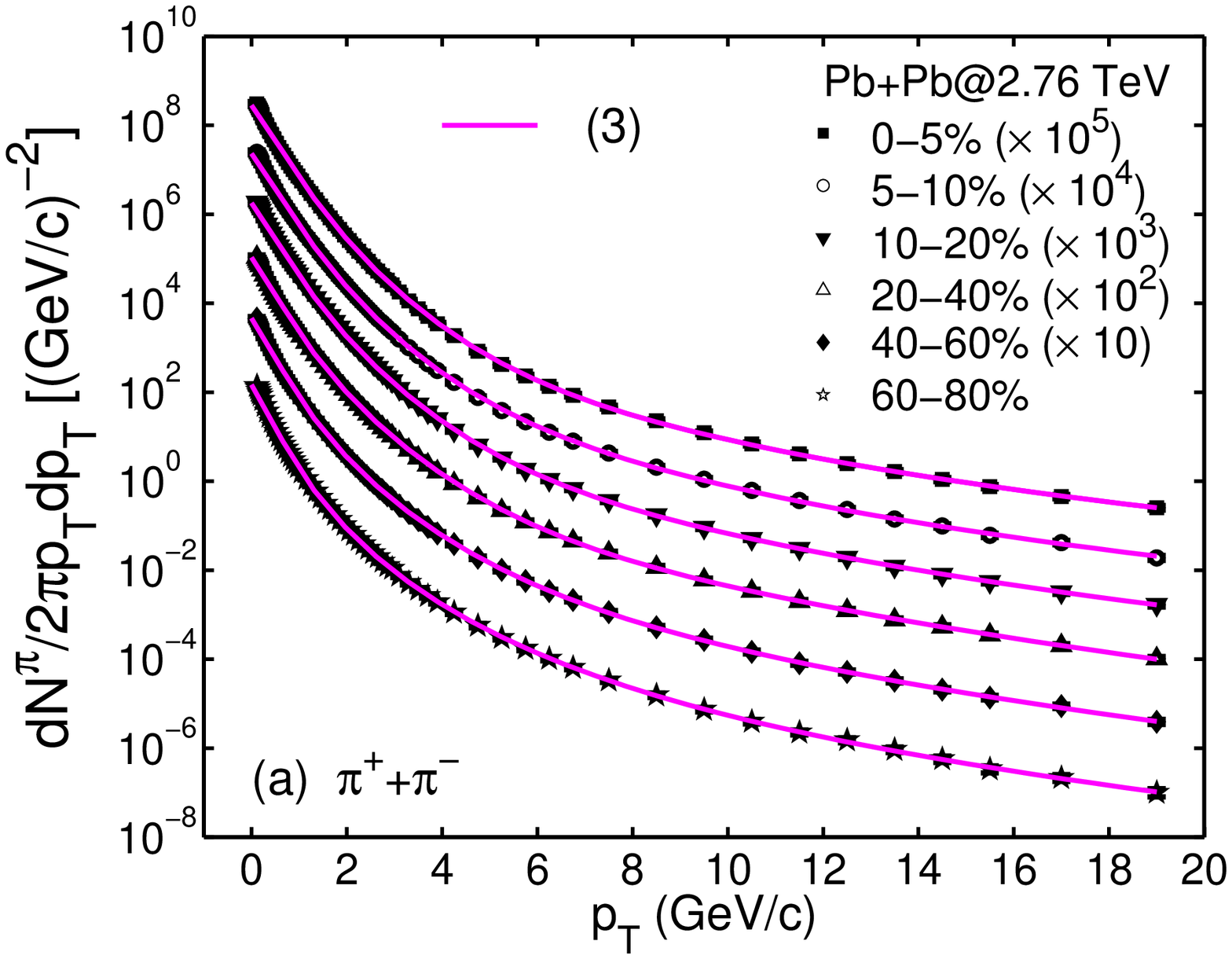}
        \includegraphics[width=0.32\textwidth]{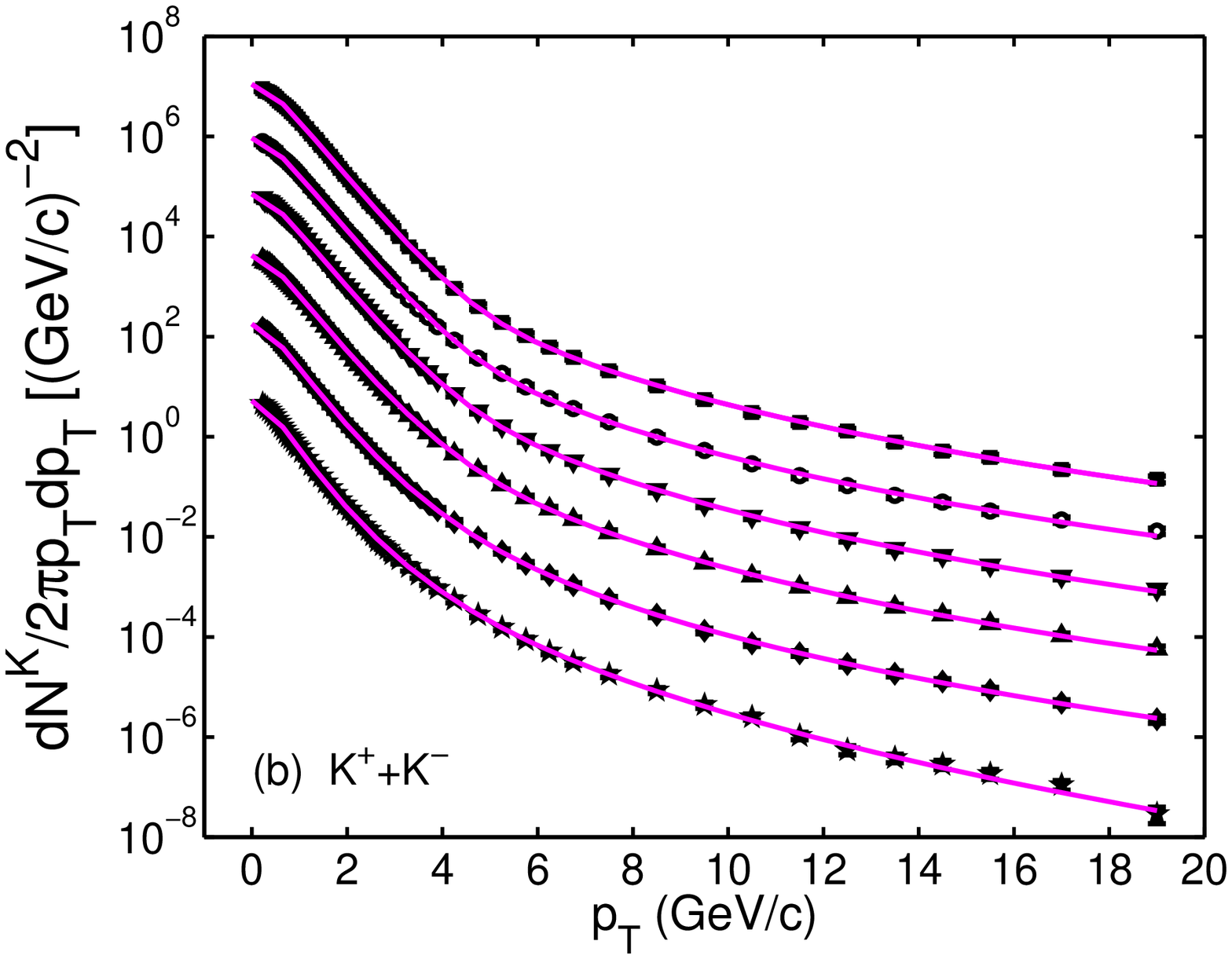}
          \includegraphics[width=0.32\textwidth]{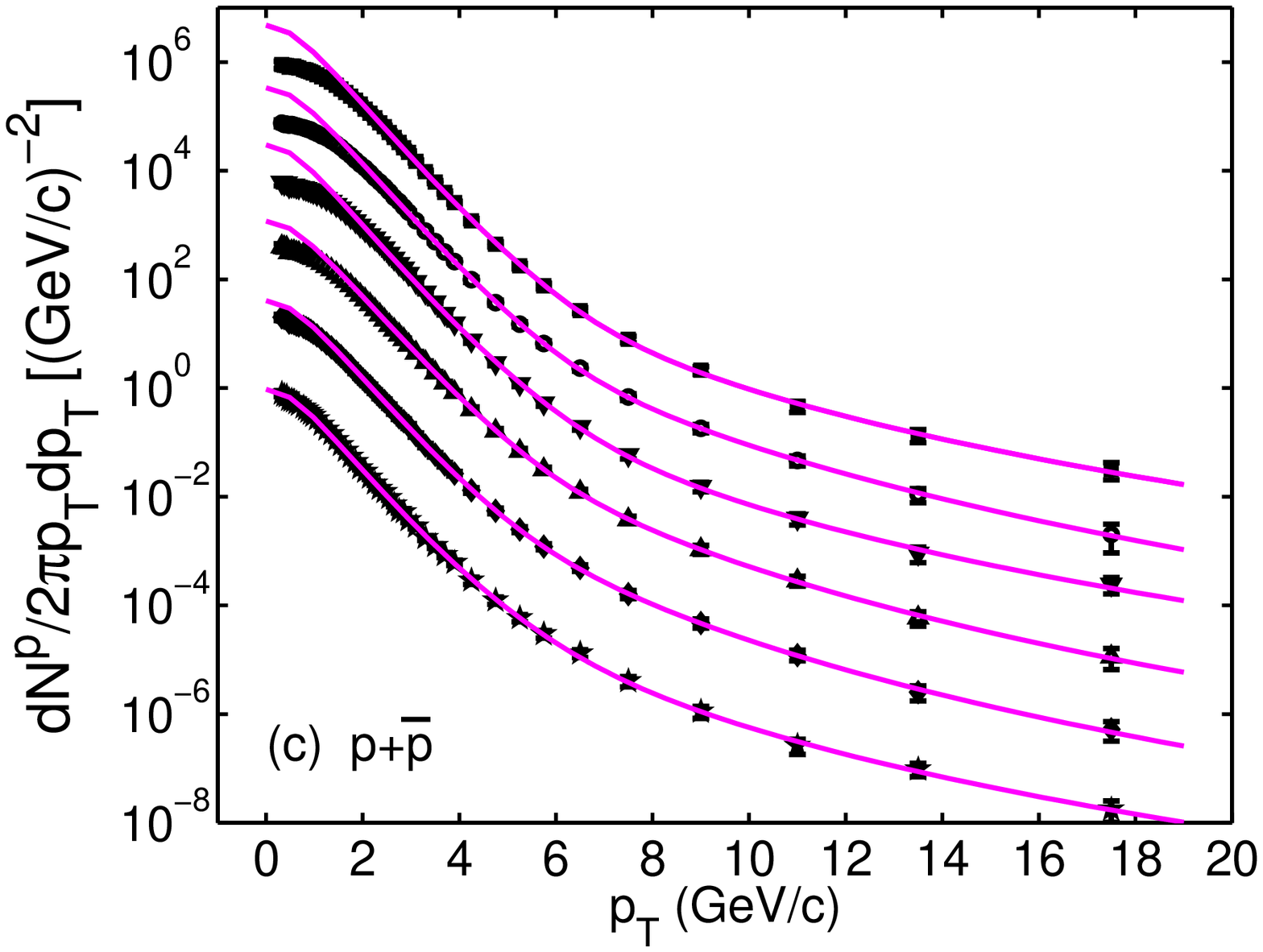}
               \end{tabular}
\caption{ (Color online) Fitting results using the double-Tsallis distribution (\ref{formula1}) for (a) $\pi^++\pi^-$, (b)  $K^++K^-$ and (c) $p+\bar p$ in Pb+Pb collisions at $\sqrt{S_{NN}}=2.76$ TeV. For a better visualization both the data and the analytical curves have been scaled by a constant as indicated. Data are taken from ALICE \cite{Adam:2015kca}.}\label{tsallis}
    \end{figure*}

As shown in Figure~\ref{tsallis}, (\ref{formula1}) reproduces the data for pions and kaons very well, while for protons the situation is different. Firstly, it is remarkable that the spectrum of protons at the centrality 60-80\% could be described by (\ref{formula1}) very well. This is reasonable and agrees with our previous work. The peripheral collisions in AA are more similar to the p+p collisions and the single Tsallis distribution can fit all the particle spectra produced in p+p collisions \cite{Zheng:2015tua}. Secondly, we also have to notice that in Figure~\ref{tsallis}(c) when $p_T\leq 2$ GeV/c, the magenta solid lines are much larger than the data for the central and less central collisions. The existence of difference is not surprising. Compared with pions and kaons, the spectra of protons demonstrate different behaviors at low $p_T$. The particle production dynamics may be different for different particles. On the one hand, a cascade particle production mechanism was proposed in p+p collisions. The heavier particles are more likely to be produced at the beginning while the light particles can be produced at all times \cite{Zheng:2015tua}. On the other hand, in the quark recombination models \cite{vg, rf, hy1}, mesons are formed by combining a quark and an antiquark while baryons by three quarks. Because different numbers of (anti)quarks participate in forming the particles, the structures of their spectra must be different. In this sense, our investigation results urge more studies on particle production mechanisms. 

\section{The  generalized Fokker-Planck solution}\label{Fokker}
\label{sec:2}

The Fokker-Plank (FP) equation has very wide applications in different fields. For instance, FP equation has been solved to study the time evolution of income distributions for different classes in a country \cite{Banerjee:2010ypa}, the rapidity spectra for net proton production at RHIC, SPS and AGS \cite{Alberico:2009epja}, and the interaction of nonequilibrated heavy quarks with the quark gluon plasma expected to be formed in heavy-ion collisions at RHIC and LHC energies \cite{Svetitsky:1987gq, Das:2010tj, Moore:2004tg, Das:2009vy}. The general form of Fokker-Planck equation is \cite{Svetitsky:1987gq},
\begin{eqnarray}
\frac{\partial P(r,t)}{\partial t}=\frac{\partial }{\partial r}\left[A(r)P(r,t) \right]+\frac{\partial^2 }{\partial r^2}\left[B(r)P(r,t)\right].
\label{formula7} 
\end{eqnarray}
The coefficients $A(r)$ and $B(r)$ are the drift and diffusion terms, respectively.  Here, $r$ represents the variable studied. With the same consideration as in (\ref{formula1}), we choose $E_T$ as the variable. Assuming $A(E_T)=A_0+\alpha E_T$ and $B(E_T)=B_0+\beta E_T^2$, one could obtain the stationary solution $P_s(E_T)$ of (\ref{formula7})
\begin{equation}
P_s(E_T)=A\frac{e^{-\frac{b}{T}\arctan{\frac{E_T}{b}}}}{[1+(\frac{E_T}{b})^2]^c},
\label{formula8}
\end{equation}
which fulfills the condition $\partial_t P_s = 0$ and the three parameters $b=\sqrt{B_0/\beta}$, $T=B_0/A_0$ and $c=1+\alpha/{2\beta}$. When $p_T\ll 1$ or $\frac{E_T}{b}\ll1$, from \eqref{formula8}, we can get\begin{equation}
P_s(E_T)\propto e^{-\frac{E_T}{T}}.
\label{formula9}
\end{equation}
On the other hand, when $p_T\gg 1$ or $\frac{E_T}{b}\gg1$, \eqref{formula8} becomes
\begin{eqnarray}
P_s(E_T)\propto p_T^{-2c}.
\label{formula10}
\end{eqnarray}

\begin{figure*}
        \centering
        \begin{tabular}{ccc}
        \includegraphics[width=0.32\textwidth]{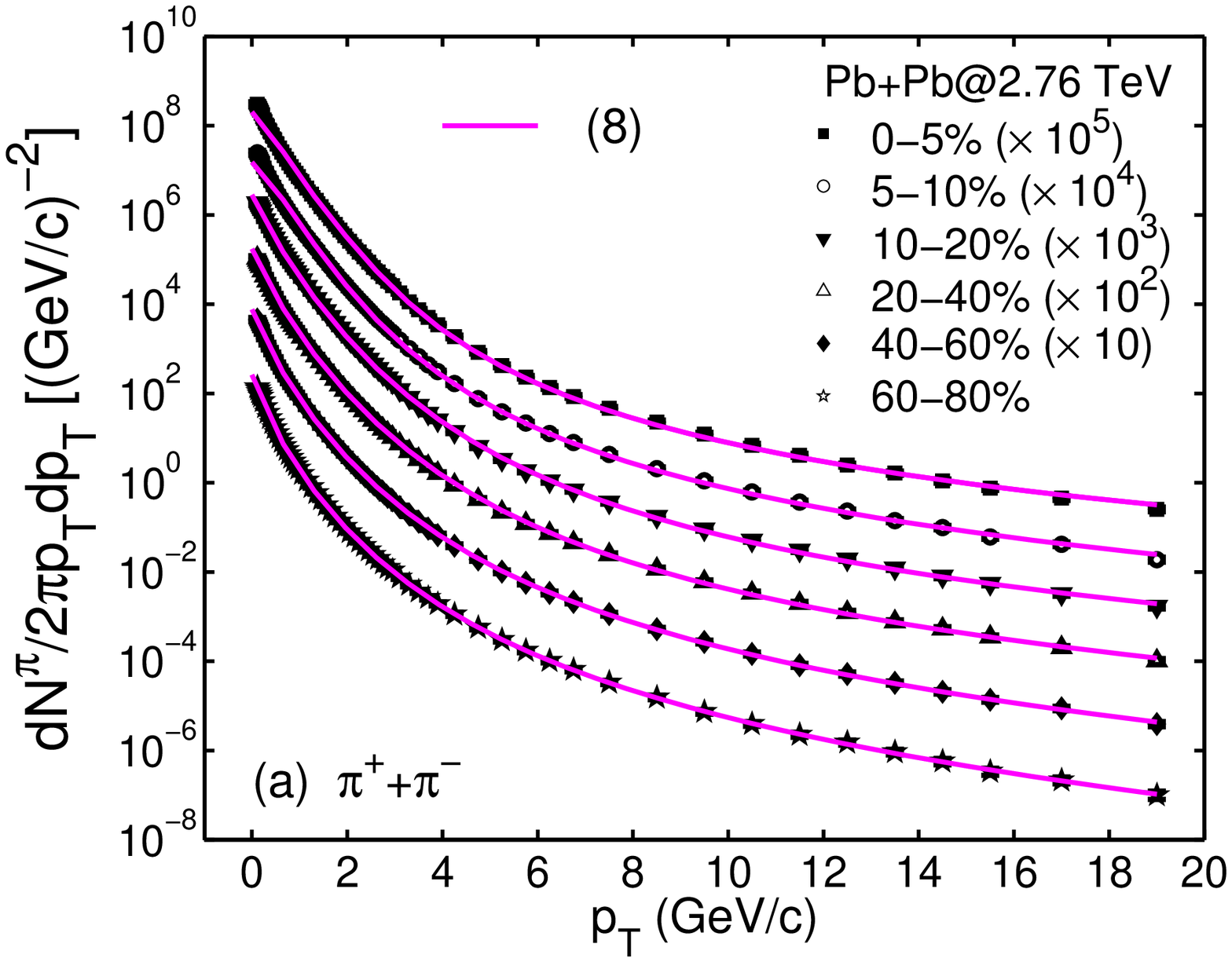}
        \includegraphics[width=0.32\textwidth]{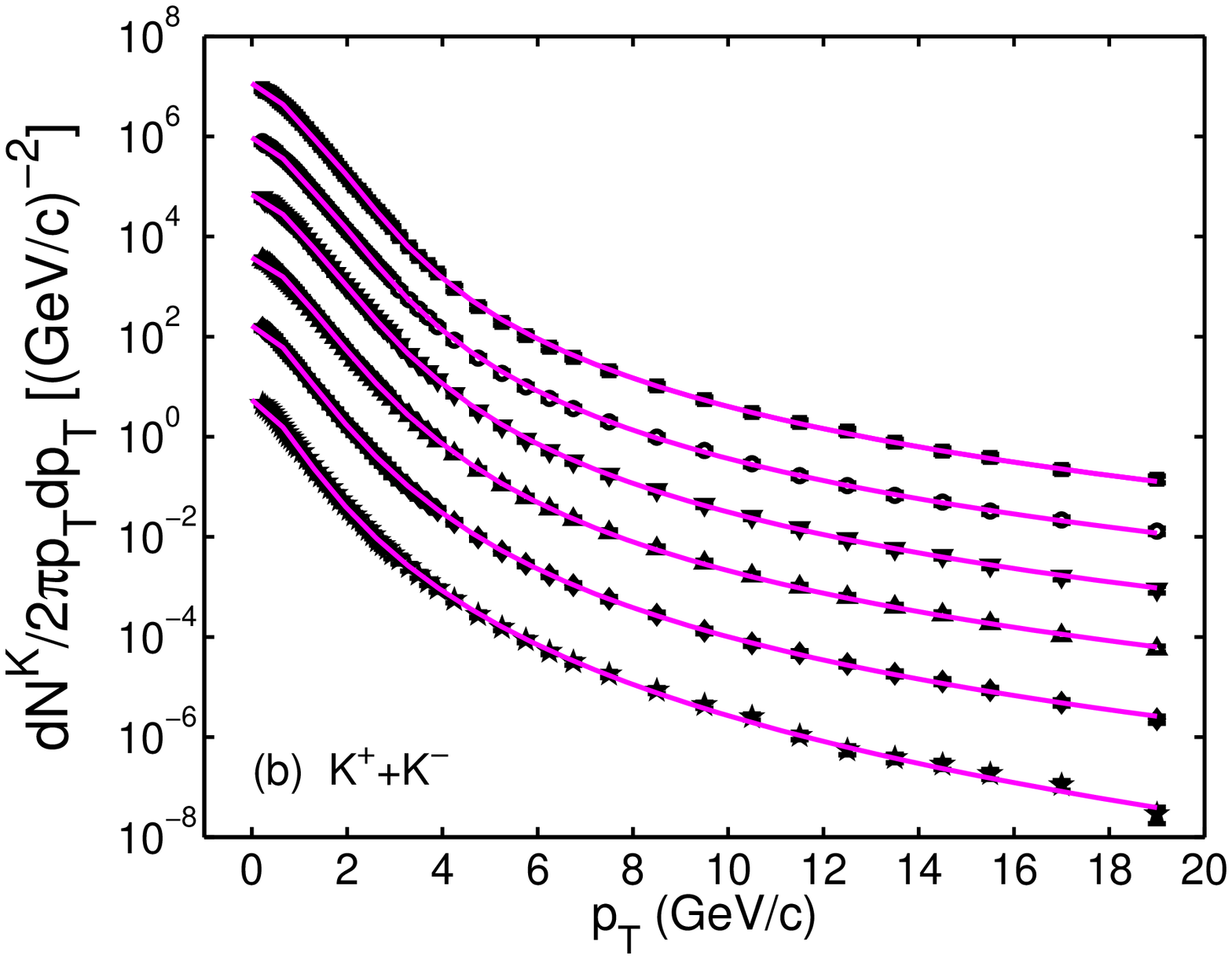}
          \includegraphics[width=0.32\textwidth]{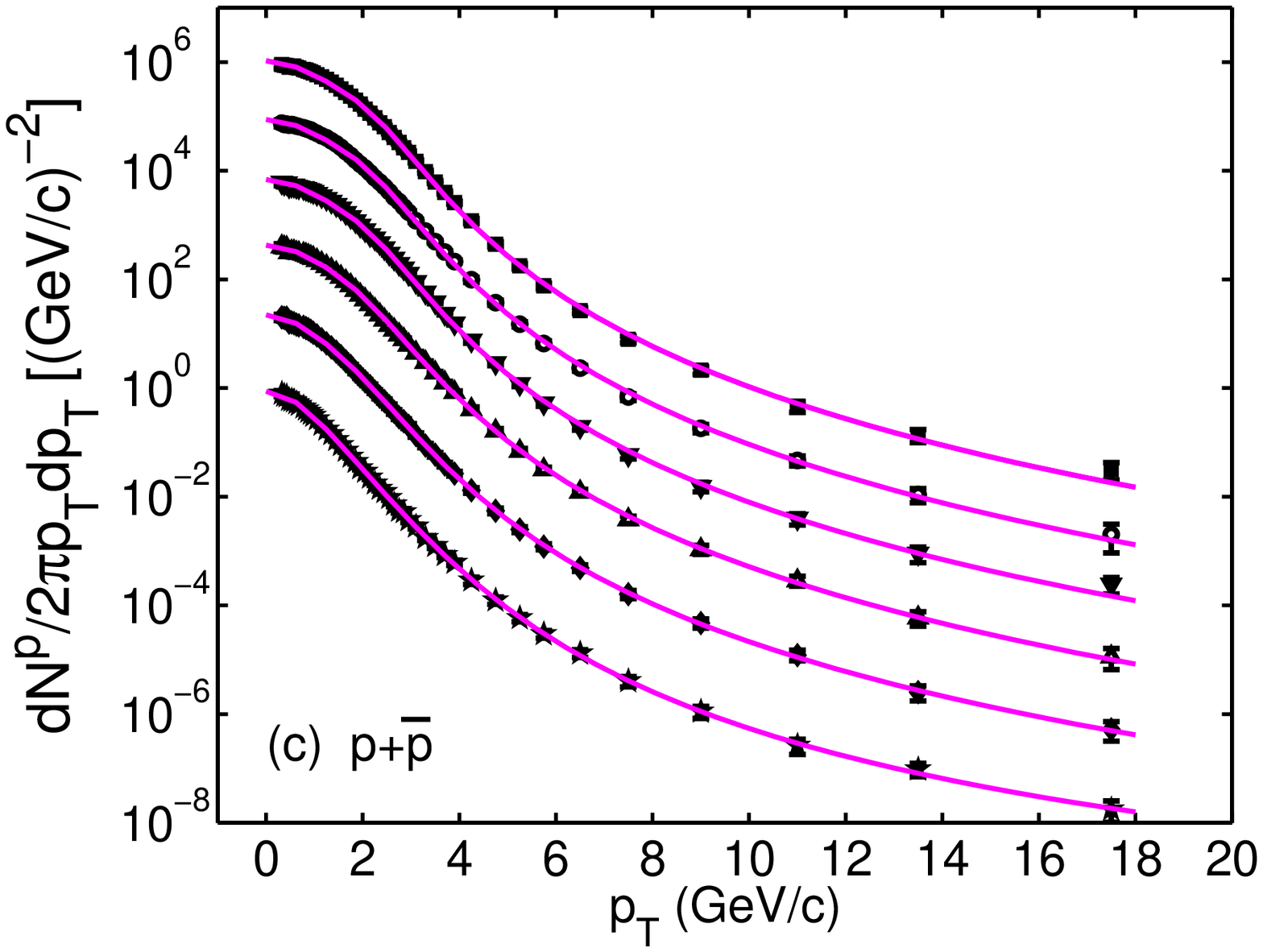}
               \end{tabular}
\caption{(Color online) Fitting results using the modified Fokker-Plank solution (\ref{formula2}) for (a) $\pi^++\pi^-$, (b)  $K^++K^-$ and (c) $p+\bar p$ in Pb+Pb collisions at $\sqrt{S_{NN}}=2.76$ TeV. For a better visualization both the data and the analytical curves have been scaled by a constant as indicated. Data are taken from ALICE \cite{Adam:2015kca}.}\label{tsallis1}
    \end{figure*}

The asymptotic behaviors of $P_s(E_T)$ are consistent with those of transverse momentum distribution of the final particle in heavy-ion collisions, which exhibits for large $p_T$ roughly a power-law distribution, whereas it becomes purely exponential for small $p_T$.  Actually, in Ref. \cite{Zheng:2015gaa}, we have proposed a formula similar to (\ref{formula8}) to fit all the particle spectra at central collisions in Pb+Pb at $\sqrt{S_{NN}}=2.76$ TeV but the power of $\frac{E_T}{b}$ in (\ref{formula8}) was fixed and changed from 2 to 4. We realize that we need to generalize the formula proposed in Ref. \cite{Zheng:2015gaa} in order to describe the spectra of identified particles at both central and non-central collisions. The generalized form of the solution of Fokker-Plank equation is
\begin{eqnarray}
E\frac{\mathrm{d}^3N}{\mathrm{d}p^3}=
A\frac{e^{-\frac{b}{T}\arctan{\frac{E_T}{b}}}}{[1+(\frac{E_T}{b})^d]^c}.
\label{formula2}
\end{eqnarray}
There are five parameters $A$, $b$, $c$, $d$ and kinetic temperature T. In Figure~\ref{tsallis1}, we redo the fitting with \eqref{formula2}. 
We can see that the fits are better than the ones with \eqref{formula1}, especially for protons. But for pions, the solid lines are a little bit lower than the data at  $p_T<1$ GeV/c for central collisions. In log scale the difference is invisible. We will show this in linear scale at Section \ref{results}. This leads us to figure out another method to illustrate the spectra of pions at very low $p_T$ region. At the same time, it also should be able to reproduce kaons and protons production. In other words, we hope that there is a universal distribution, which could describe the spectra of the different final particles for the whole $p_T$ region.

         \begin{figure*}
        \centering
        \begin{tabular}{ccc}
        \includegraphics[width=0.32\textwidth]{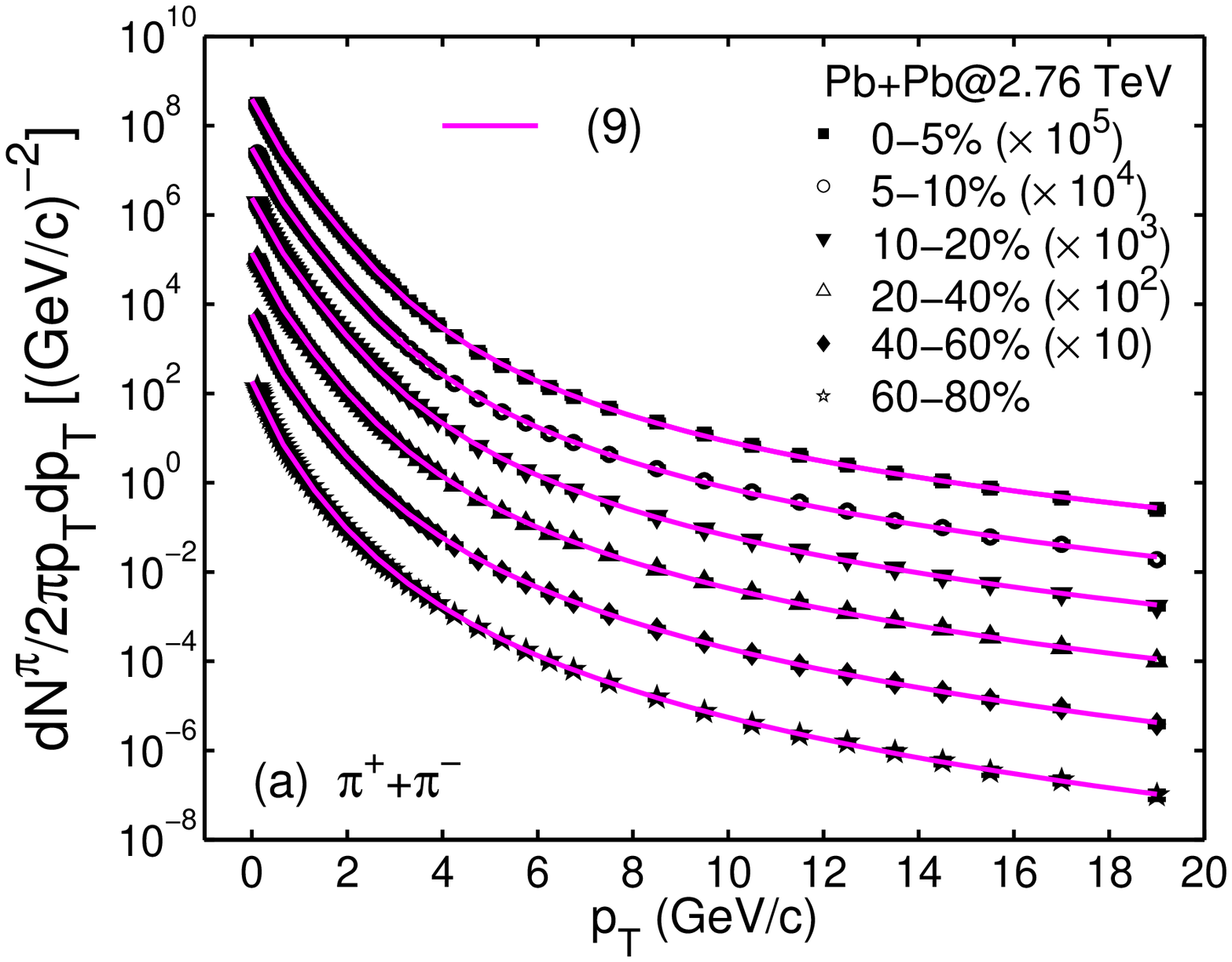}
      \includegraphics[width=0.32\textwidth]{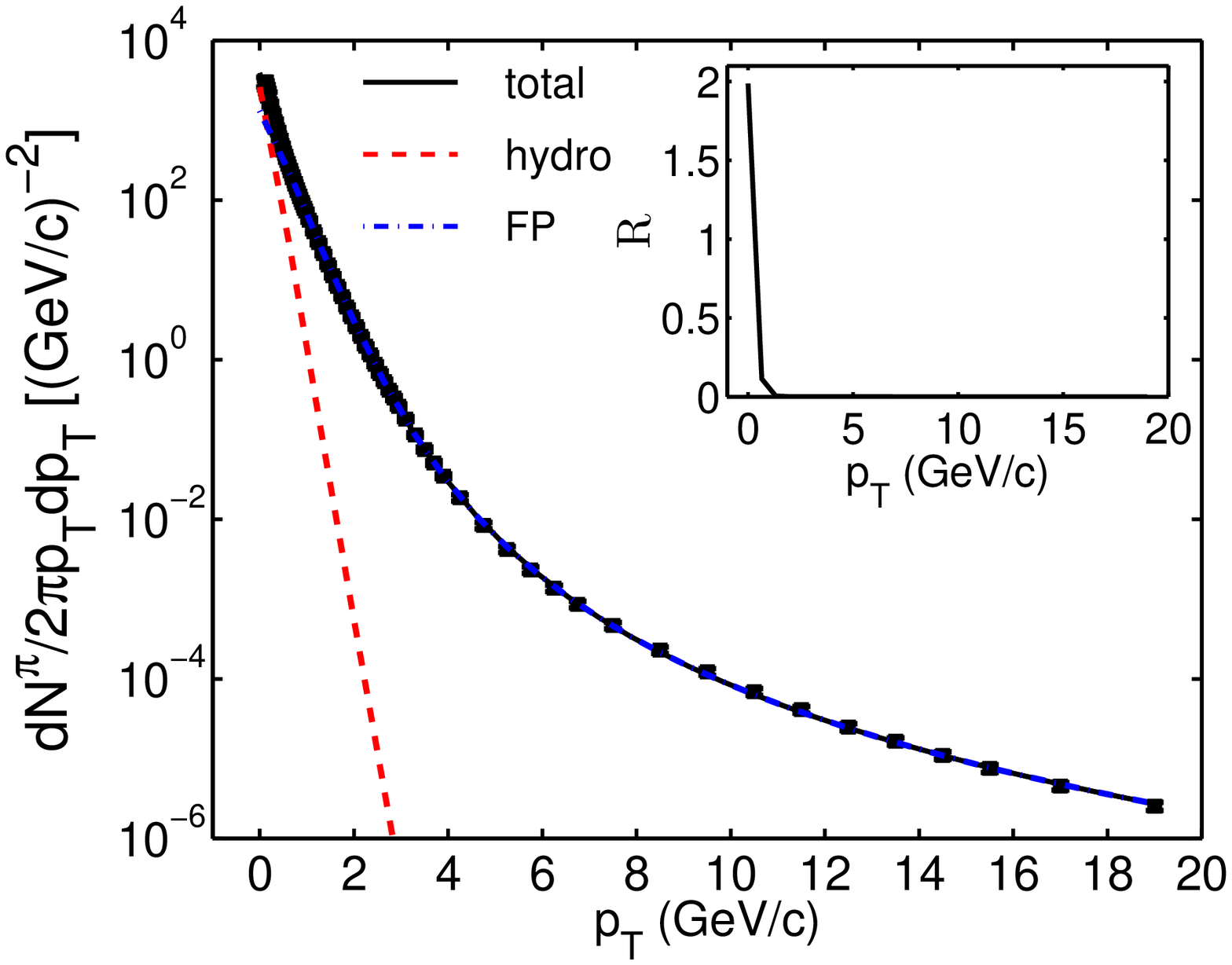}
               \end{tabular}
\caption{(Color online) Fitting results using the new two-component distribution (\ref{formula4}) for (a) $\pi^++\pi^-$, while in (b) the red (dashed) and blue (dash-dotted) lines show the hydrodynamic term (hydro) and the generalized Fokker-Plank term (FP), respectively. The inset is for the ratio of the two terms, see the context. Data are taken from ALICE \cite{Adam:2015kca}. }\label{hy+fokker}
    \end{figure*}
         
\section{A new two-component distribution}\label{hydro}
Recognizing that both the double-Tallis distribution and the generalized Fokker-Plank solution are not able to fully reproduce the observed structure at low $p_T$ region, it is argued that one should realize that the soft and hard particles have different production mechanisms. We need to figure out how to make the fit for pions better at very low $p_T$ since \eqref{formula2} can fit kaons and protons very well,  see Figure~\ref{tsallis1}(b) and \ref{tsallis1}(c). The bulk of low-$p_T$ particles originates from the "quark-gluon soup" formed in the heavy-ion collision and has an exponential distribution, while the high-$p_T$ tail accounts for the mini-jets that pass through the hot medium, a process that can be described in pQCD. When a large colliding system is formed, one should also take the effects of the "collective motion" into account \cite{Schnedermann:1993ws}. Thus, in heavy-ion collisions multiparticle production is usually considered in terms of relativistic hydrodynamics, contrary to the widely used thermodynamic approaches for pp, $\gamma$p and $\gamma\gamma$ collisions \cite{Tsallis:1987eu}.  Therefore, we introduce a hydrodynamic extension to the generalized Fokker-Plank solution \eqref{formula2} to improve the fit for pions in Figure \ref{tsallis1}(a). 

As shown in Figure \ref{hy+fokker}(a), we present an example of the use of this method for pions at  six different centralities in Pb+Pb collisions using the following formula
\begin{eqnarray}
\frac{\mathrm{d}N}{p_T \mathrm{d}p_T}=A_e \int_{0}^{R}r\mathrm{d}rm_TI_0(\frac{p_T\sinh\rho}{T_e})
K_1(\frac{m_T\cosh\rho}{T_e})+
A\frac{e^{-\frac{b}{T}\arctan{\frac{E_T}{b}}}}{[1+(\frac{E_T}{b})^d]^c}.
\label{formula4}
\end{eqnarray}
The results are indeed encouraging, even though two more parameters $A_e$ and $T_e$ are added. The solid lines include the total contributions from the hydrodynamic and generalized Fokker-Plank terms in \eqref{formula4}. To get a clear picture of the two terms, we show them in different color lines for 0-5\% in  Figure \ref{hy+fokker}(b). To show the difference clearly a ratio $\mathcal{R}$ of the hydrodynamic term over the generalized Fokker-Plank term is plotted in the inset of Figure \ref{hy+fokker}(b) as a function of $p_T$. It indicates that the generalized Fokker-Plank term is absolutely dominated at $p_T$>1 GeV/c, while the contribution from the hydrodynamic term definitely can not be ignored at very low $p_T$ region. When the contribution from hydrodynamic term is extremely small, \eqref{formula4} is same as \eqref{formula2} and it can reproduce the particle spectra for kaons and protons.

         \begin{figure*}
        \centering
        \begin{tabular}{ccc}
        \includegraphics[width=0.32\textwidth]{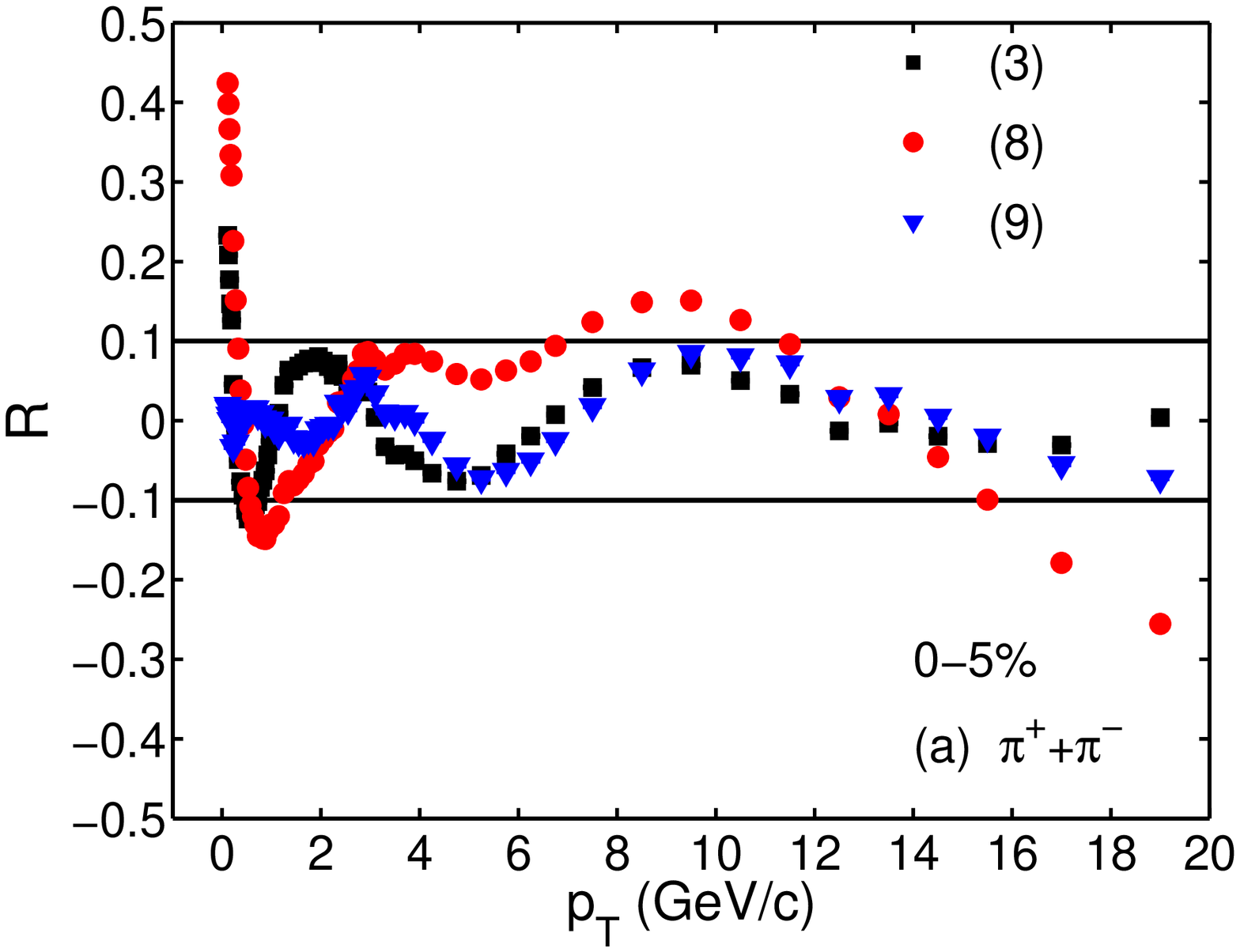}
        \includegraphics[width=0.32\textwidth]{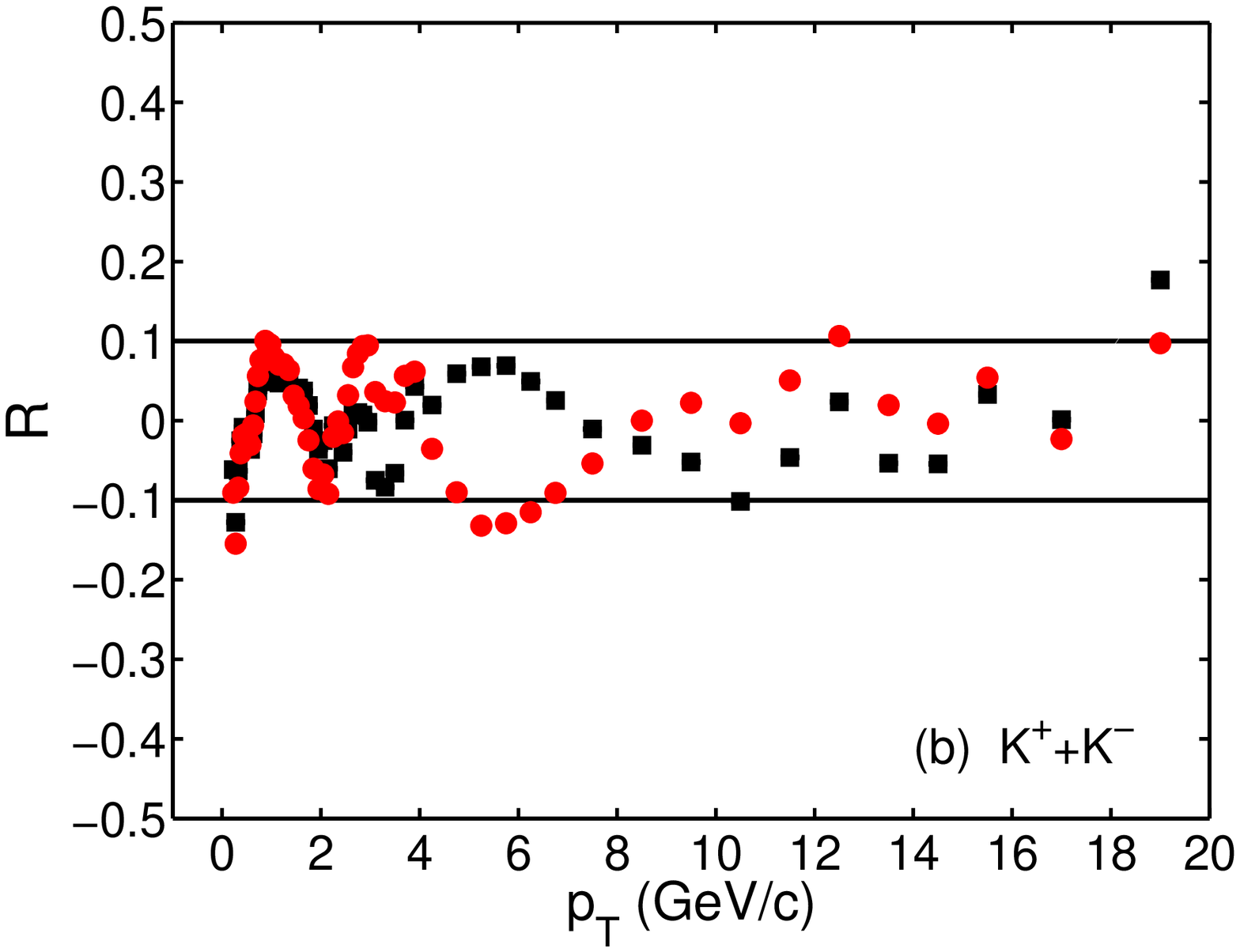}
          \includegraphics[width=0.32\textwidth]{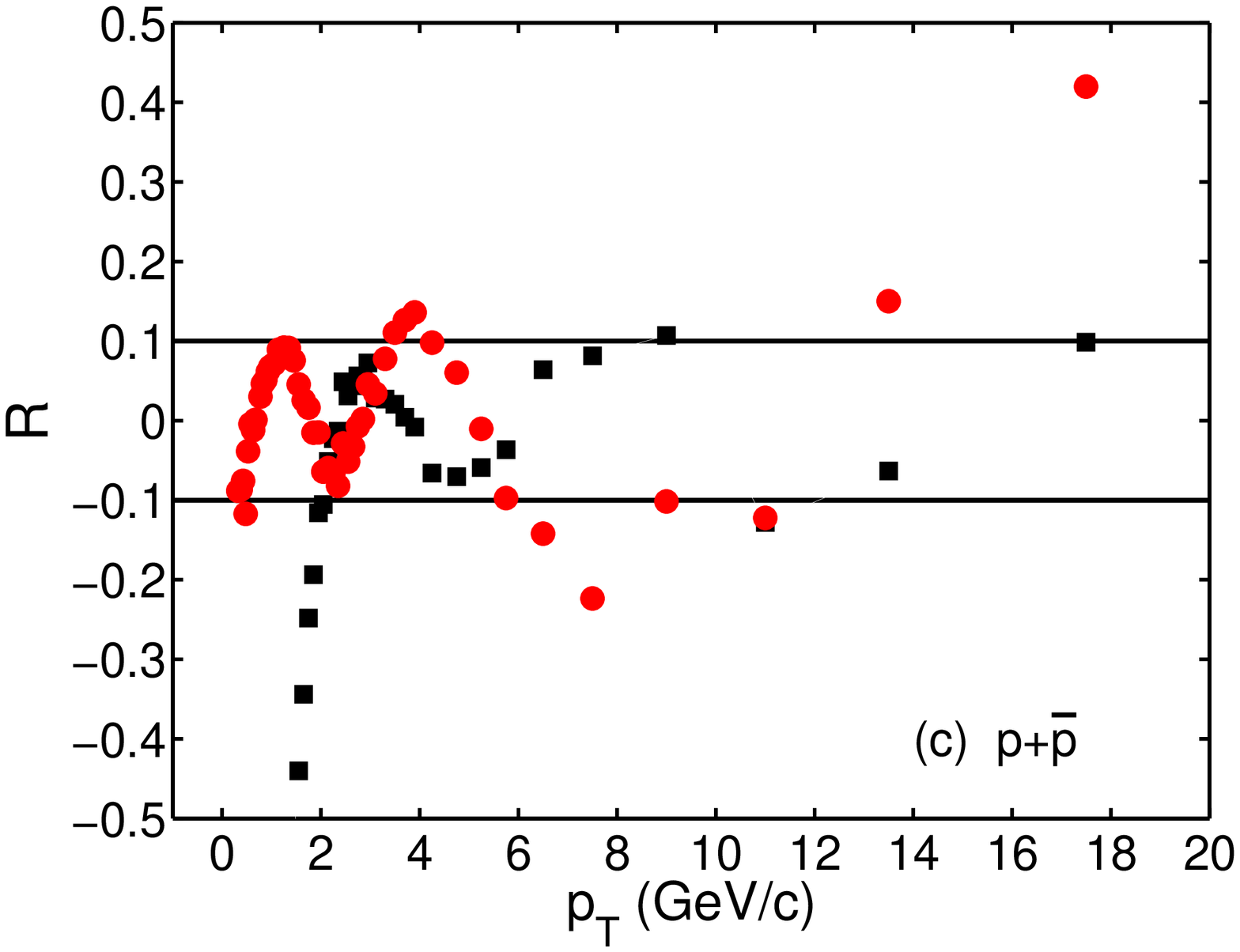}
               \end{tabular}
\caption{(Color online) The relative discrepancies of (\ref{formula1}), (\ref{formula2}) and (\ref{formula4}) from the $p_T$ spectra for (a) $\pi^++\pi^-$, (b)  $K^++K^-$ and (c) $p+\bar p$ at 0-5\% shown in Figures~\ref{tsallis}, \ref{tsallis1} and \ref{hy+fokker}, respectively.}\label{ratio1}
    \end{figure*}
         \begin{figure*}
        \centering
        \begin{tabular}{ccc}
        \includegraphics[width=0.32\textwidth]{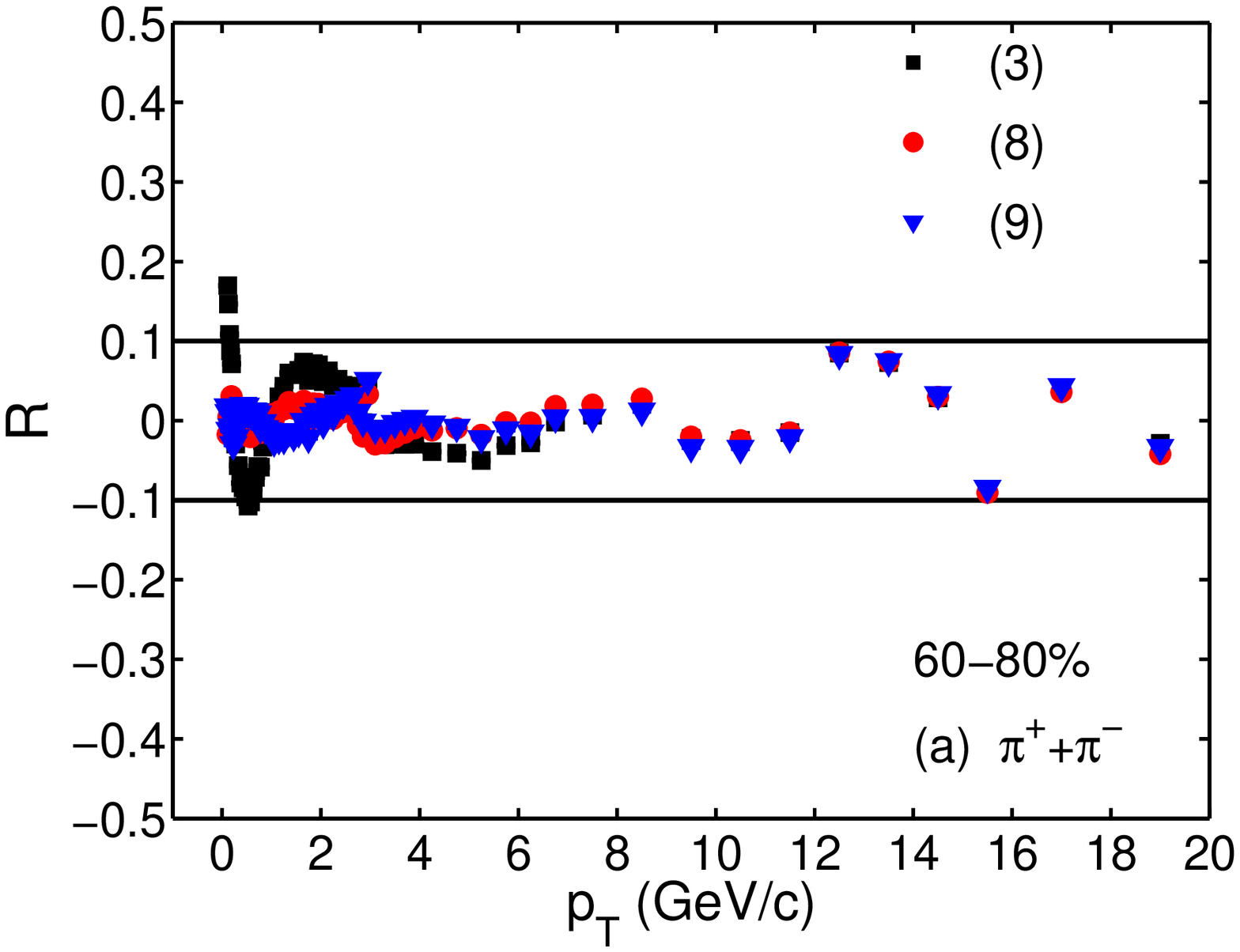}
        \includegraphics[width=0.32\textwidth]{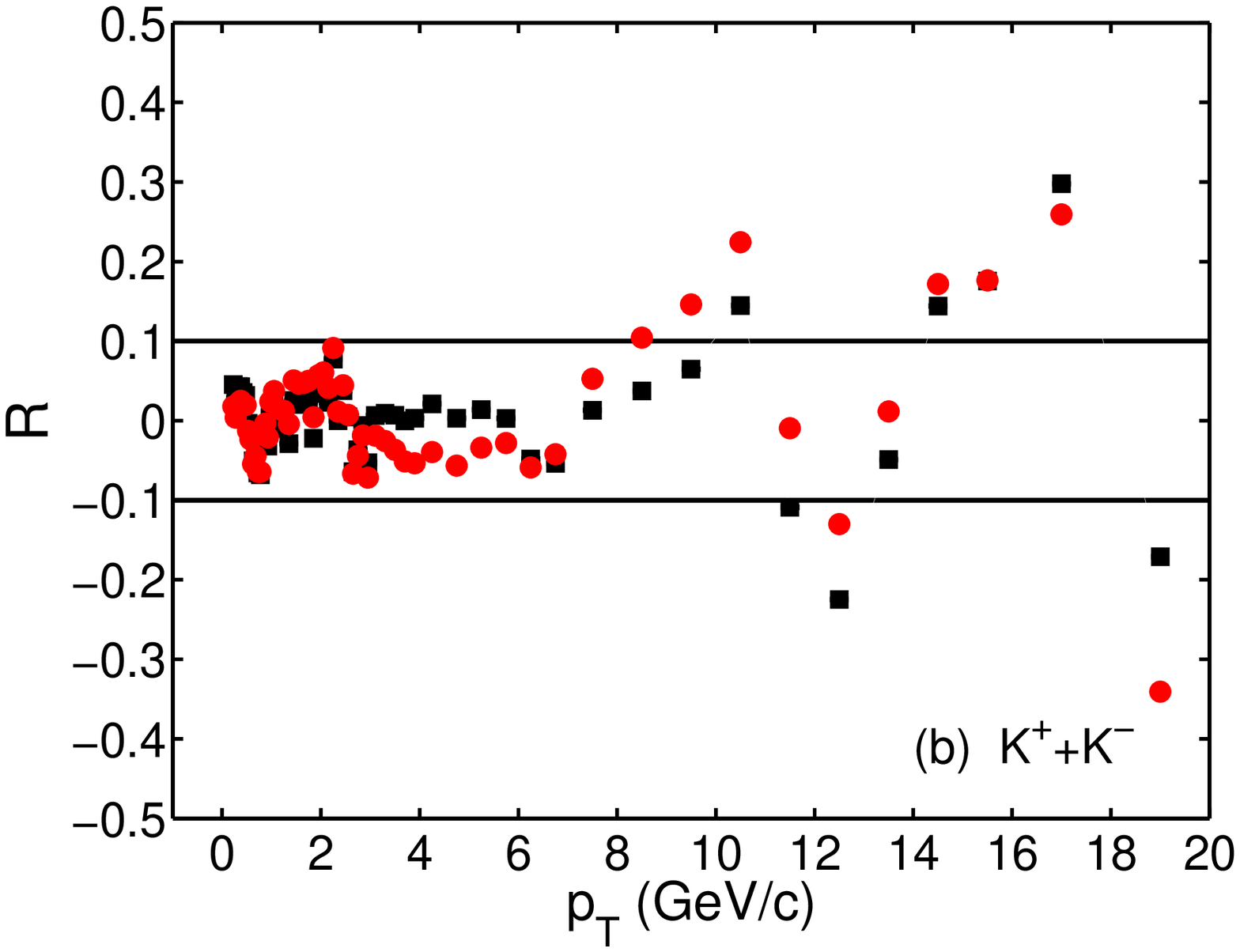}
          \includegraphics[width=0.32\textwidth]{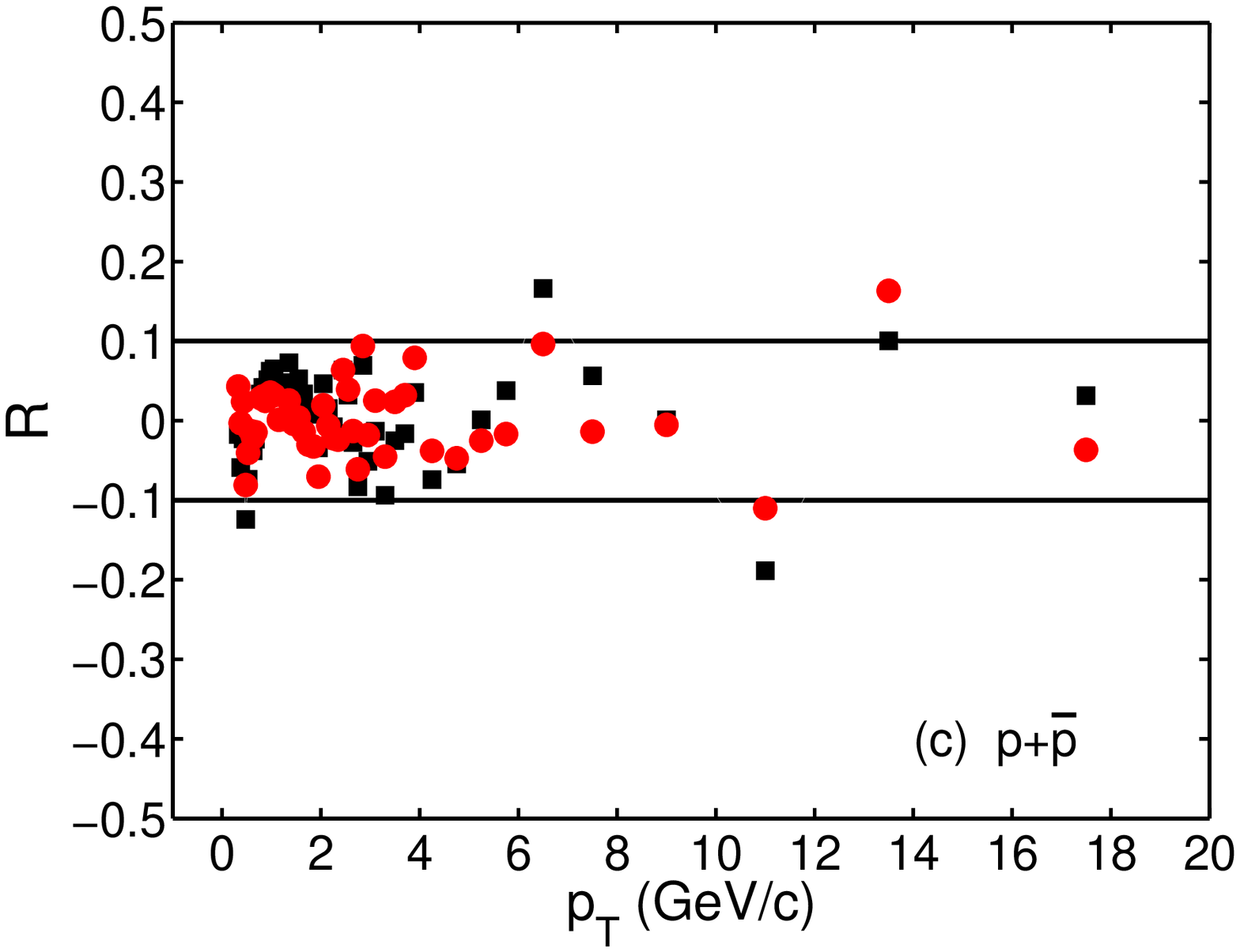}
               \end{tabular}
\caption{(Color online) The relative discrepancies of (\ref{formula1}), (\ref{formula2}) and (\ref{formula4}) from the $p_T$ spectra for (a) $\pi^++\pi^-$, (b)  $K^++K^-$ and (c) $p+\bar p$ at 60-80\% shown in Figures~\ref{tsallis}, \ref{tsallis1} and \ref{hy+fokker}, respectively.}\label{ratio2}
    \end{figure*}
 \section{Discussion}\label{results}
Generally speaking, the results shown in the previous sections indicate that the three different methods, which are the double-Tsallis distribution \eqref{formula1},  the generalized Fokker-Plank solution \eqref{formula2} and the new two-component distribution \eqref{formula4}, can describe the experimental data. But we want to understand which distribution is the optimal choice.  To have a clearer picture, we evaluate the degree of agreement of the fitted results with the experimental data. One can calculate the ratio between the experimental data and the fitted results, which is defined as
$$R=(data-fitted)/data.$$

Figure \ref{ratio1} shows the ratio $R$, calculated by  (\ref{formula1}), (\ref{formula2}) and (\ref{formula4}) respectively, as a function of the transverse momentum $p_T$ in linear scale for the centrality 0-5\%.  In Figure \ref{ratio1}(a), one can see that all points for pions produced in most central collisions from (\ref{formula4}) are in the range from -0.1 to 0.1, while the relative discrepancies from \eqref{formula1} and (\ref{formula2}) are large at low $p_T$ region. For kaons, (\ref{formula1}) and (\ref{formula2}) have the similar fitting power as shown in Figure \ref{ratio1}(b), the deviation of the fitting results from data is less than 10\%.  While, for protons, Figure \ref{ratio1}(c)  establishes that (\ref{formula1}) is not good for low $p_T$ region, which can be easily seen from Figure \ref{tsallis}(c).

For the sake of the comprehensive comparison, we should also check the relative discrepancies of the three equations at other centralities. Here, we only plot the results for the centrality 60-80\%. Except a few points, the relative discrepancies of the three equations for pions, kaons and protons are in good agreement with the data with deviation from the data less than 10\%. Remarkably, the fluctuations for pions, kaons, and protons are much smaller than those for the centrality 0-5\%. In other words, the three distributions agree with each other to describe better the particle spectra produced in peripheral collisions, which are similar to p+p collisions. 

Based on these analyses, we can conclude that (\ref{formula4}) is the best one among the three distributions, which is composed of a hydrodynamic term and the generalized Fokker-Plank solution. It could well describe the spectra from central to peripheral collisions for pions, kaons and protons. The proposed hydrodynamic extension (\ref{formula4}) of \eqref{formula2} slightly modifies the description of the experimental data for pions at low $p_T$ region, which also gives insight to the particle production mechanism. 
               
\section{summary}\label{summary}
In this paper, we have made a detailed study of the double-Tsallis distribution, the generalized Fokker-Plank solution and the new two-component distribution, by fitting the transverse momentum spectra of pions, kaons and protons in Pb+Pb collisions at $\sqrt{S_{NN}}=2.76$ TeV. The double-Tsallis distribution can fit the particle spectra except the big deviation observed for proton at $p_T<2$ GeV/c for central and less central collisions, while the generalized Fokker-Plank solution is not able to describe the spectra of pions at very low $p_T$. Therefore, we propose a new two-component distribution as a hydrodynamic extension of the generalized Fokker-Plank solution accounting for the collective motion effect in order to fit all the particle spectra in Pb+Pb collisions, especially for extremely low $p_T$ region. According to these results, we can conclude that the new two-component distribution is the optimal method. From these analyses, we get more information about the particle production mechanism in Pb+Pb collision. We also wish more exciting results could be found in Pb+Pb collisions at $\sqrt{S_{NN}}=5.02$ TeV.

\section*{Conflict of Interests}
The authors declare that there is no conflict of interests regarding the publication of this article.

\section*{Acknowledgments}
This work is supported by the NSFC of China under Grant No.\ 11205106.

\end{document}